\documentclass{aa}  
\usepackage[draft]{todonotes}   

\usepackage{comment}
\usepackage{graphicx}
\usepackage{txfonts}
\begin{document}

   \title{Refractory elements in the gas phase for comet 67P/Churyumov-Gerasimenko}

   \subtitle{Possible release of atomic Na, Si, and Fe from nanograins}

   \author{Martin Rubin \inst{1}
          \and
          Kathrin Altwegg \inst{1}
          \and
          Jean-Jacques Berthelier \inst{2}
          \and
          Michael R. Combi \inst{3}
          \and
          Johan De Keyser \inst{4}
          \and
          Frederik Dhooghe \inst{4}
          \and
          Stephen Fuselier \inst{5,6}
          \and
          Tamas I. Gombosi \inst{3}
          \and
          Nora Hänni \inst{1}
          \and
          Daniel Müller \inst{1}
          \and
          Boris Pestoni \inst{1}
          \and
          Susanne F. Wampfler \inst{7}
          \and 
          Peter Wurz \inst{1}
           }

   \institute{
             Physikalisches Institut, University of Bern, Sidlerstrasse 5, CH-3012 Bern, Switzerland\\
             \email{martin.rubin@unibe.ch}
           \and
             Laboratoire Atmosphères, Milieux, Observations Spatiales, Institut Pierre Simon Laplace, CNRS,
             Université Pierre et Marie Curie, 4 Avenue de Neptune, F-94100 Saint-Maur, France
           \and
             Department of Climate and Space Sciences and Engineering, University of Michigan, 2455 Hayward, Ann Arbor, MI 48109, USA
           \and
             Royal Belgian Institute for Space Aeronomy, BIRA-IASB, Ringlaan 3, B-1180 Brussels, Belgium
           \and
             Space Science Directorate, Southwest Research Institute, 6220 Culebra Rd., San Antonio, TX 78228, USA
           \and
             Department of Physics and Astronomy, The University of Texas at San Antonio, San Antonio, TX 78249, USA
           \and
             Center for Space and Habitability, University of Bern, Gesellschaftsstrasse 6, CH-3012 Bern, Switzerland
             }

   \date{Received 13 September 2021 / Accepted 16 December 2021}

 
  \abstract
   {Gas-phase sodium, silicon, potassium, and calcium were previously identified in mass spectra recorded in the coma of comet 67P/Churyumov-Gerasimenko, the target of the  European Space Agency's Rosetta mission. The major release process for these atoms was identified as sputtering by the solar wind. More recently, remote observations of numerous comets over a range in heliocentric distances revealed the presence of metal atoms of iron and nickel that had been released either from the nucleus or from a distributed source with a short scale length. Sputtering, however, has been dismissed as a major release process due to the attenuation of the solar wind in the comae of some of the observed targets.}
   {We investigated the presence of refractory species in the gas phase of the coma of 67P/Churyumov-Gerasimenko. This investigation includes a period close to perihelion when the solar wind was likely absent from the near-nucleus region due to the increased cometary activity. Additionally, we extended our search to iron and nickel.}
   {We analyzed in situ data from the Rosetta/ROSINA Double Focusing Mass Spectrometer DFMS.}
   {We found that gas-phase silicon was present throughout the Rosetta mission. Furthermore, the presence of sodium and iron atoms near the comet's perihelion confirms that sputtering cannot be the sole release process for refractory elements into the gas phase. Nickel was found to be below the detection limit. The search for parent species of any of the identified gas phase refractories has not been successful. Upper limits for a suite of possible fragment species (SiH, SiC, NaH, etc.) of larger parent and daughter species have been obtained. Furthermore, Si did not exhibit the same drop in signal as do common cometary gases when the spacecraft is pointed away from the nucleus. The combined results suggest that a direct release of elemental species from small grains on the surface of the nucleus or from small grains in the surrounding coma is a more likely explanation than the previous assumption of release via the dissociation of gaseous parent molecules.}
   {}


   \maketitle
%

\section{Introduction}
Comet 67P/Churyumov-Gerasimenko (hereafter \mbox{67P/C-G}) has a pronounced bi-lobate shape. Together with its elliptical orbit \citep{Maquet2015} and the orientation of the spin axis \citep{Sierks2015}, strong seasonal variations in the composition of the coma volatiles have been observed \citep{Hassig2015}. The major species identified in the coma of \mbox{67P/C-G} were H$_2$O, CO$_2$, CO, and O$_2$ \citep{LeRoy2015,Bieler2015}. However, there were not only volatile species that were found -- in addition, refractory elements Na, Si, K, and Ca in the gas phase were reported early on in the Rosetta mission \citep{Wurz2015} from the Rosetta Orbiter Spectrometer for Ion and Neutral Analysis \citep[hereafter ROSINA,][]{Balsiger2007}.  \cite{Wurz2015} attributed the release of these species to sputtering by solar wind. One reason behind this notion was that their abundances were co-correlated but anti-correlated to the major gases in the coma. The total gas densities from the less active southern hemisphere early in the mission were consistent with a scenario of unhindered access on the part of the solar wind to the nucleus. The higher volatile densities above the simultaneously more active northern hemisphere resulted in a collisional attenuation of solar wind ion energies before reaching the surface. Furthermore, the derived relative abundances of Na/Si, K/Si, and Ca/Si were in agreement \citep{Rubin2020} with the measured ratios in the cometary dust obtained by the COmetary Secondary Ion Mass Analyzer \citep[COSIMA;][]{Kissel2007,Bardyn2017}. The Si signal early in the mission, when Rosetta was close to the nucleus of \mbox{67P/C-G}, was sufficient to obtain the $^{29}$Si/$^{28}$Si and $^{30}$Si/$^{28}$Si isotope ratios in the comet \citep{Rubin2017}. Both the $^{29}$Si and $^{30}$Si heavy isotopes were lower in abundance with respect to $^{28}$Si when compared to the corresponding solar ratios.

Refractory elements are well known to be present in the cometary dust phase. Examples include the Stardust samples returned from comet Wild 2 \citep{Zolensky2006} and the dust collected in the coma of comet \mbox{67P/C-G} analyzed by the COSIMA instrument \citep{Bardyn2017}. 

However, the same elements have also been identified in the gas phase in other comets. Iron atoms, for instance, have been observed in the tail of comet C/2006 P1 (McNaught) at perihelion \citep{Fulle2007}. More recently, abundant amounts of Fe and Ni atoms were detected in the coma of about 20~comets, with some of them  far away from the Sun \citep{Manfroid2021,Bromley2021}. Furthermore, atomic Ni was observed in the interstellar object \mbox{2I/Borisov} \citep{Guzik2021}. Up to that point, the corresponding emission lines of metallic atoms had only been observed in comets \citep{Jones2018} or exocomets \citep{Kiefer2019} that were either colliding with or passing close by the Sun or their host star, respectively. The presence of refractory atoms in cometary comae over a range of heliocentric distances from $<$0.2 to $>$3~au \citep{Manfroid2021} sets constraints on the release processes, making these elements a key target of study.

\section{Rosetta mission}
The European Space Agency launched the Rosetta mission on 2~March 2004. The target was Jupiter-family comet \mbox{67P/C-G} whose current orbit is the result of a close encounter with Jupiter in 1959 \citep{Maquet2015}. Given its orbital period of almost 6.5~years and its perihelion distance of 1.24~au, comet \mbox{67P/C-G} has already crossed the inner Solar System several times. Rosetta arrived at its target in early August 2014 and followed the comet for more than two years along its journey around the Sun. Together with its lander Philae, Rosetta carried out an in-depth investigation of the comet's nucleus and surrounding coma. Rosetta carried a complement of 11~instruments and another 10~instruments on the lander Philae. Over the two years, \mbox{67P/C-G} covered a range in heliocentric distance, from $\sim$3.6~au, to perihelion (which occurred on 13 August 2015), and out again to almost 4~au. The mission concluded at the end of September 2016, when the Rosetta spacecraft landed on the comet's surface.

\section{\label{sec:DFMS}ROSINA DFMS -- Instrument description}
Among the suite of instruments was ROSINA \citep{Balsiger2007}, which consisted of two mass spectrometers, RTOF (Reflectron-type Time-Of-Flight) and DFMS (Double Focusing Mass Spectrometer), as well as the COmet Pressure Sensor (COPS). As a double-focusing mass spectrometer in Nier-Johnson configuration \citep{Balsiger2007}, DFMS  has a mass resolution of m/$\Delta$m~=~3000 at the 1\% level on mass/charge~=~28~Da/e. Neutral gas entering the instrument was ionized inside the source by electron impact using 45~eV electrons. If the primary neutral contains two or more atoms, other fragment ions could also form, such as: CO$_2$~+~e$^-$~$\rightarrow$~CO$_2^+$~+~2~e$^-$ and CO$_2$~+~e$^-$~$\rightarrow$~CO$^+$~+~2~e$^-$. Hence, DFMS measured ions and fragment ions originating from the neutrals entering the ion source. Ions from the surrounding coma, however, did not have appropriate energies to pass through the system due to the bias voltage of DFMS' ion source. Most ions formed from the entering neutrals were singly charged and hence we refrained from denoting the charge state in the following, except for the subset of doubly charged ions, for instance, CO$_2^{++}$.

The newly formed ions were accelerated and guided through a narrow, 14~$\mu$m slit before being deflected by 90 degrees in a toroidal electrostatic analyzer (ESA) with energy resolution of $\Delta$E/E~$\sim$~1\%. Following the ESA, a 60~degree deflection occurred in the field of a permanent magnet. Given the adjustable ion energy and electrostatic fields in combination with the permanent magnet, the instrument was tuned such that only ions with a suitable mass/charge (m/Q) ratio made it through the analyzer section. This work is focused on the relevant high-resolution data. In this mode, the mass resolution was increased through a series of two electric quadrupole fields acting as a zoom lens before the ion beam finally impinged on the micro channel plate (MCP) detector. In the MCP detector, the primary ion beam was converted into an amplified shower of secondary electrons which were then collected on a position-sensitive Linear Electron Detector Array (LEDA) with two rows (A and B) of 512 pixels each \citep{Nevejans2002}. For heavier species with m/Q ratios $>$70~Da/e an additional post acceleration potential of 1000~V in front of the MCP increased the detection efficiency.

A single MCP spectrum covered a range of m/Q around a central m/Q, for instance, $\pm$0.25~Da/e at m/Q~=~28~Da/e. Individual mass peaks in DFMS are fitted using the sum of two Gaussians, one with approximately 90\% of the signal amplitude and the second one with about 10\% of the amplitude and about 3 times the width of the first one. More details about DFMS peak shapes can be found in \cite{DeKeyser2019a,DeKeyser2019b}. The high-resolution mass scans analyzed in this paper were performed by applying a suitable set of voltages to obtain a mass spectrum around each integer mass and $\sim$20~s signal accumulation per spectrum. Following each measurement, a new set of voltages was applied including a settling time of 10~s before acquiring the next mass spectrum. DFMS covered the m/Q range from 12 to 180~Da/e, although most measurement sequences did not cover the full range. A standard m/Q scan with the MCP detector in the range of 13 to 100~Da/e took roughly 45~minutes. These non-contemporaneous measurements introduced additional uncertainties, for instance, changes in the cometary environment during the time of a full mass scan. However, it permitted a change in the amplification of the MCP detector from one spectrum to the next. As a result, the dynamic range of the instrument was increased up to 10$^{10}$ (whereas each individual spectrum was limited to a dynamic range of $\sim$10$^{4}$).

   \begin{figure*}
   \centering
   \includegraphics[width=0.7\textwidth]{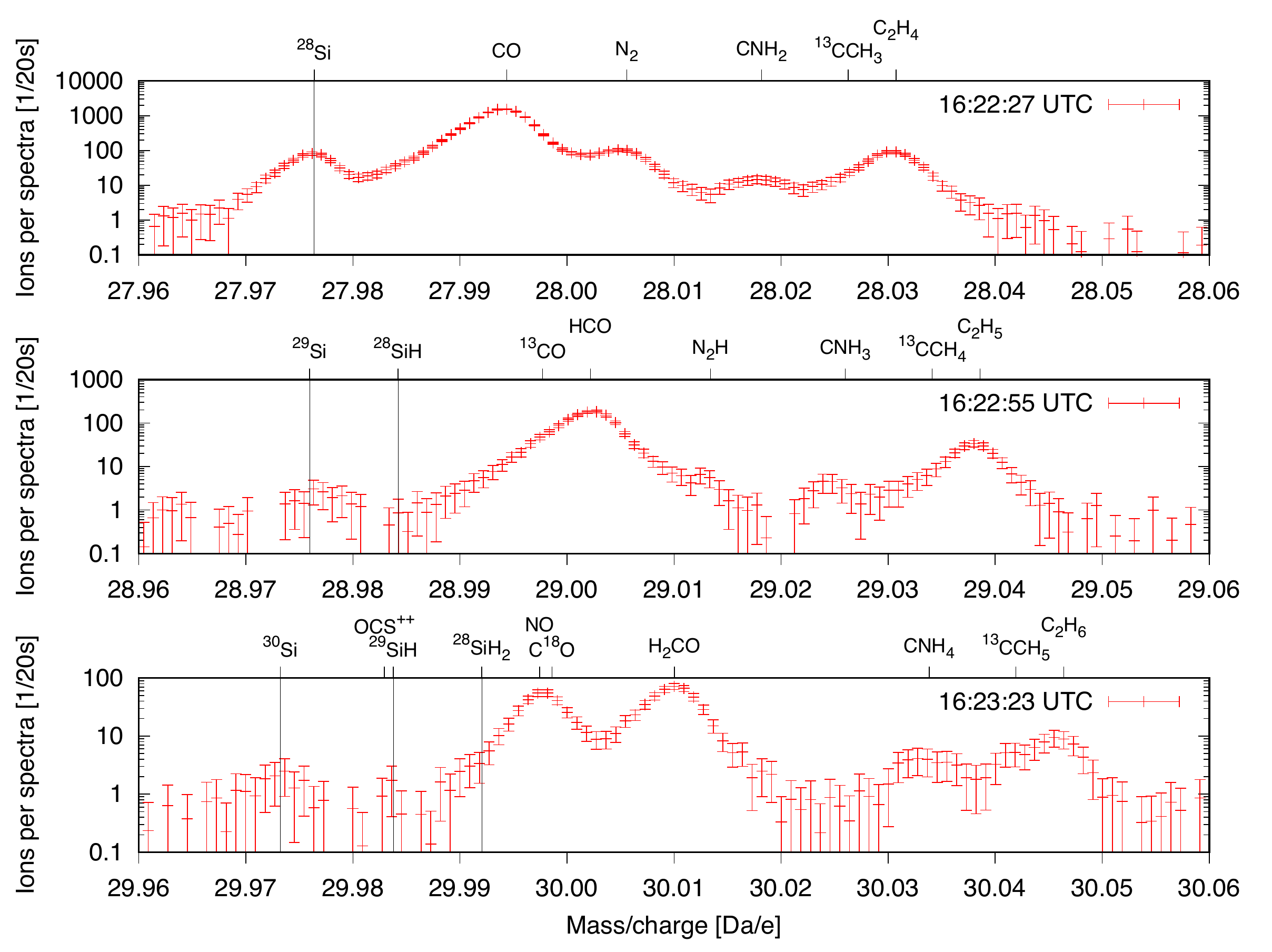}
   \caption{ROSINA DFMS m/Q~=~28 to 30~Da/e spectra in log scale top to bottom, measured during the slew on 31 July 2015, 16:22:27 (ESA Planetary Science Archive (PSA) filename and DFMS detector row: MC\_20150731\_162059567\_M0222.TAB, row~B), 16:22:55 (MC\_20150731\_162127567\_M0222.TAB, row~B), and 16:23:23~UTC (MC\_20150731\_162155567\_M0222.TAB, row~B) at off-nadir angles decreasing from 150 to 148~degrees (see Fig.~\ref{fig:GreatCircleScan}). Indicated on the x-axis on top is the m/Q of a suite of species. Vertical lines mark the exact masses of species of interest (parents and fragments), i.e., $^{28}$Si with the minor isotopes $^{29}$Si and $^{30}$Si as well as where $^{28}$SiH, $^{29}$SiH, and $^{28}$SiH$_2$ would be expected. We note that except for $^{28}$Si, the top labels do not include the mass number of the major isotope of an element, e.g., $^{12}$C, $^{14}$N, $^{16}$O, etc.}
              \label{fig:m28m29m30}%
    \end{figure*}

\section{\label{sec:results}Results}
Analysis of the m/Q~=~28~Da/e spectra in the coma of comet \mbox{67P/C-G} revealed the presence of Si in the gaseous phase throughout the mission, covering the full range of heliocentric distances from 1.24~au to almost 4~au. Furthermore, Na has been identified and occasional signals of Fe have been obtained near perihelion. Atomic Ni, on the other hand, was below the detection limit. This section summarizes these results and the conditions under which they were obtained. 

\subsection{\label{sec:Si}Silicon (Si)}
The stable isotopes of Si are found at m/Q~=~28–30~Da/e and an example set of spectra is seen in Fig.~\ref{fig:m28m29m30}. The three spectra were obtained back-to-back on 31 July 2015 at a cometocentric distance of $\sim$200~km and a heliocentric distance of 1.25~au. In the top panel, the major isotope $^{28}$Si, indentified by the black vertical line, is clearly visible in the left shoulder of CO, a major volatile in the coma of \mbox{67P/C-G} (for nominal coma molecules we refer to the major isotopologue and refrain from indicating the mass number, i.e., CO instead of $^{12}$C$^{16}$O). In this spectrum, the $^{28}$Si signal is comparable to that of N$_2$ and even higher than some of the organic species and fragments on the same mass line -- for instance, C$_2$H$_4$. The reason for this high relative amplitude is that the spectra were obtained when Rosetta and DFMS were pointing away from the nucleus. These details are discussed later in this paper. In the lower two panels, the low signals of the two minor isotopes, $^{29}$Si and $^{30}$Si, are shown \citep[$^{29}$Si/$^{28}$Si~=~0.0434$\pm$0.0050 and $^{30}$Si/$^{28}$Si~=~0.0263$\pm$0.0038; ][]{Rubin2017}.

There is a distinct absence of $^{28}$SiH and $^{28}$SiH$_2$ on mass per charges 29 and 30~Da/e, respectively. This absence limits the possible parent species of Si in the coma of \mbox{67P/C-G}. As an example, the NIST mass spectrum with unit mass resolution of silane SiH$_4$ \citep{Linstrom2001} shows the ($^{28}$SiH+$^{29}$Si) signal to be roughly equal to the $^{28}$Si signal. Furthermore, the ($^{28}$SiH$_2$+$^{29}$SiH+$^{30}$Si) signal is about four times larger than the $^{28}$Si signal. If we neglect small contributions from the minor isotopes $^{29}$Si and $^{30}$Si, we obtain $^{28}$SiH~$\sim$~1~$\times$~$^{28}$Si and $^{28}$SiH$_2$~$\sim$~4~$\times$~$^{28}$Si. From our data, we obtain approximate upper limits of $^{28}$SiH~$\le$~$^{28}$Si/100 and $^{28}$SiH$_2$~$\le$~$^{28}$Si/50. Silane and other Si-bearing species, which fragment into SiH or SiH$_2$ in proportions larger than the upper limits given above, are therefore excluded as parent species present near Rosetta.

As stated above, the spectra in Fig.~\ref{fig:m28m29m30} were obtained during a large slew, when the spacecraft was off-pointing from the nucleus. These conditions are of particular interest as these data provide useful information on the origin of the Si. Common volatiles found in \mbox{67P/C-G}'s coma and released from the nucleus showed a strong decrease in the signal, by almost a factor of ten, during these slews. This decrease is expected when the spacecraft is pointing away from the nucleus. Si, on the other hand, remained approximately constant. As a result, there is much less interference between common cometary volatiles and the refractory species during these slews, for instance, between CO and $^{28}$Si. As a consequence, potential parent species of Si could be investigated in more detail than possible when the spacecraft was continuously pointed at the nucleus and the resulting interferences were much more pronounced. Furthermore, the different behavior of Si with respect to common volatiles indicates a much more uniform and extended distribution of atomic Si in the coma, possibly associated with a distributed source such as charged nanograins. This possibility is discussed in Section~\ref{sec:discussion}.

   \begin{figure*}
   \centering
   \includegraphics[width=0.7\textwidth]{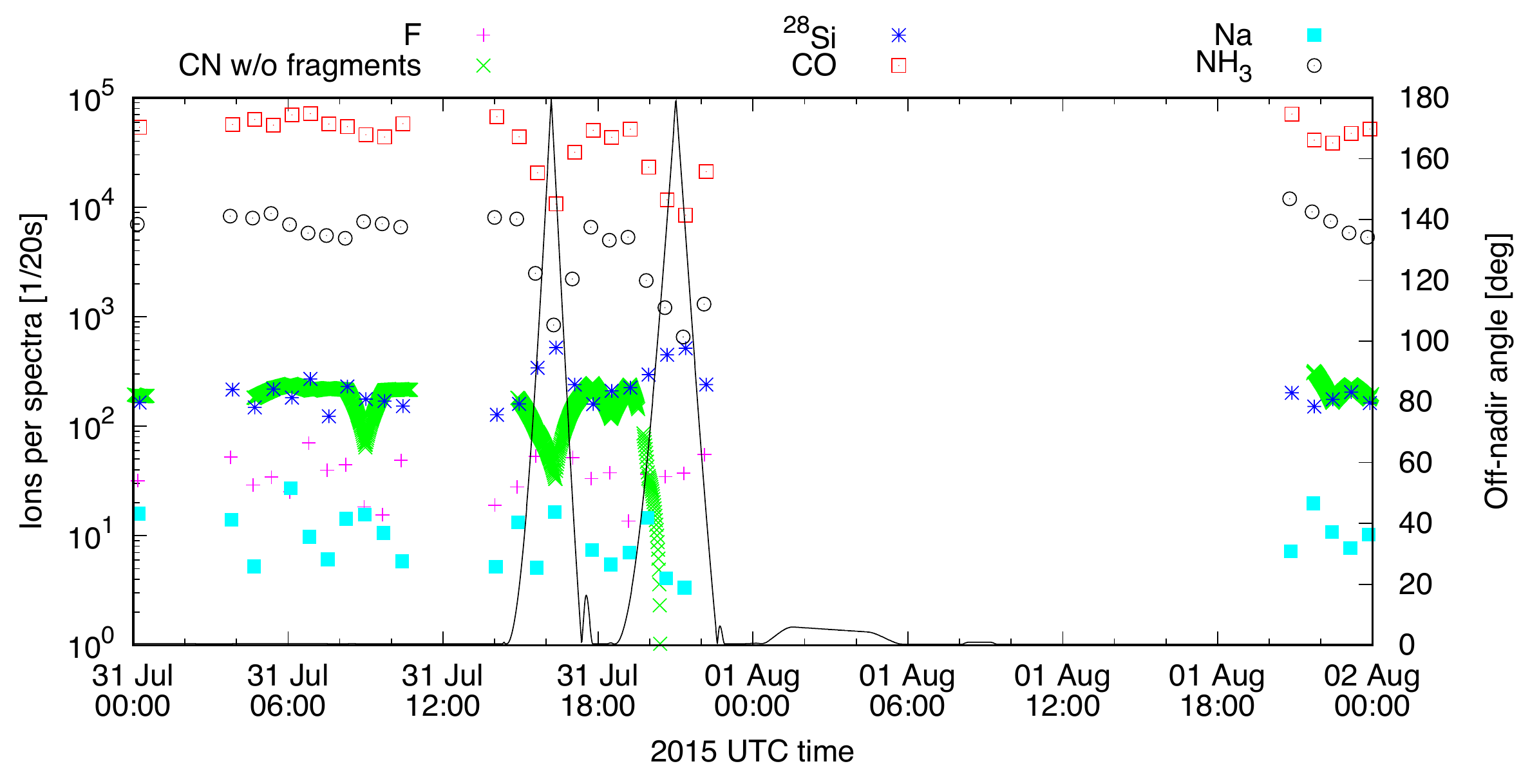}
   \caption{ROSINA DFMS timeline of CO, NH$_3$, CN, F, Na, and $^{28}$Si signals (left y-axis), measured between 31 July and 2 August 2015. Fragmentation of HCN$\rightarrow$CN has been subtracted from the CN signal. The solid black line marks the off-nadir angle (right y-axis) including the two great circle scans on 31 July 2015. The cometocentric distance was $\sim$200 km. For most of 1 August 2015 no neutral gas data is available as DFMS was operated in ion mode.}
              \label{fig:GreatCircleScan}%
    \end{figure*}

\begin{figure*}
   \centering
   \includegraphics[width=0.7\textwidth]{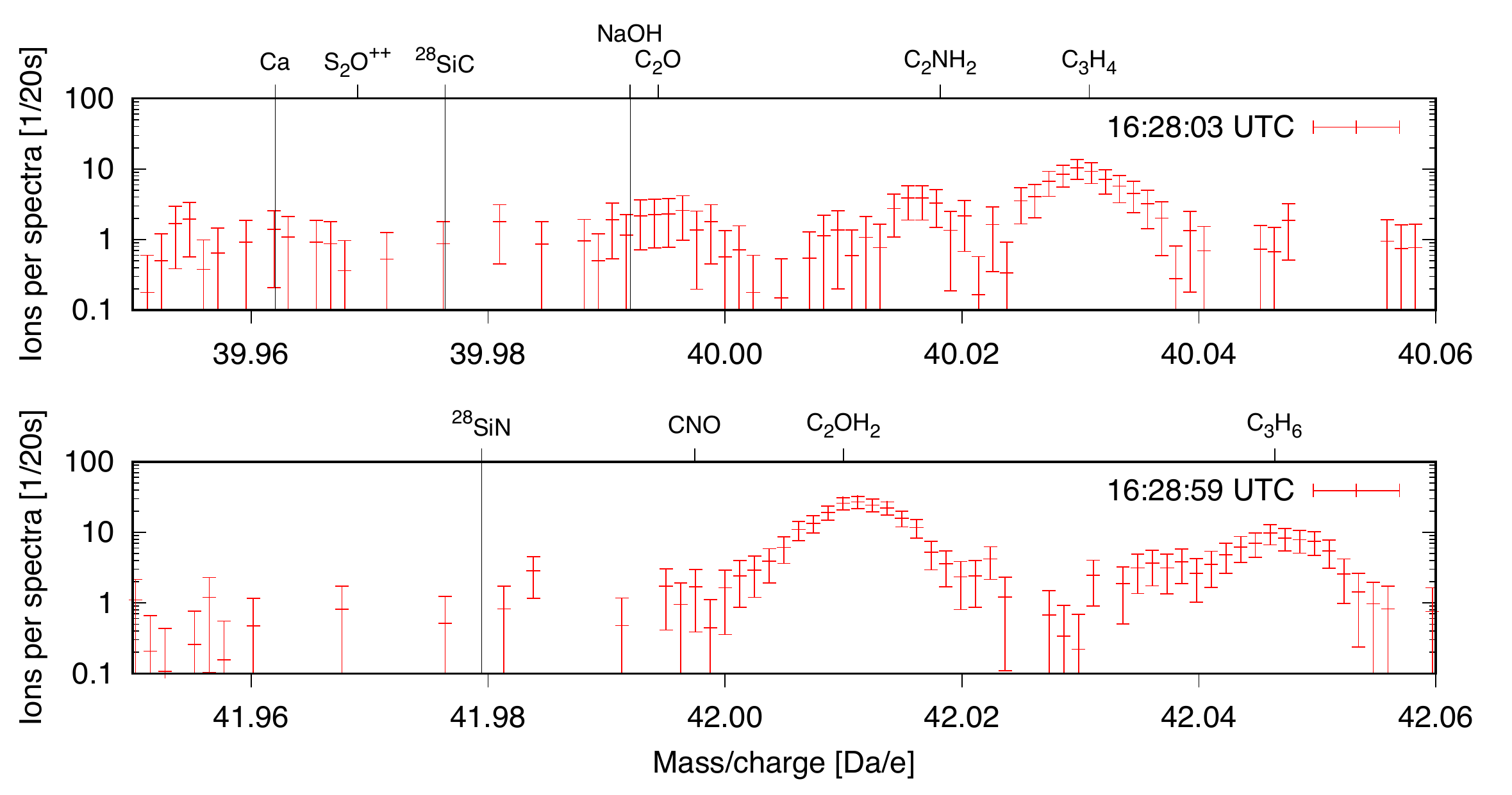}
   \caption{ROSINA DFMS m/Q~=~40 and 42~Da/e spectra top and bottom, measured on 31 July 2015, 16:28:03 (MC\_20150731\_162635570\_M0222.TAB, row~B) and 16:28:59~UTC (MC\_20150731\_162731570\_M0222.TAB, row~B) during the slew (off-nadir angles 135 and 132~degrees, see Fig.~\ref{fig:GreatCircleScan}). Indicated are the locations where Ca, $^{28}$SiC, and $^{28}$SiN would be expected.}
              \label{fig:m40m42}%
    \end{figure*}

\begin{figure*}
   \centering
   \includegraphics[width=0.7\textwidth]{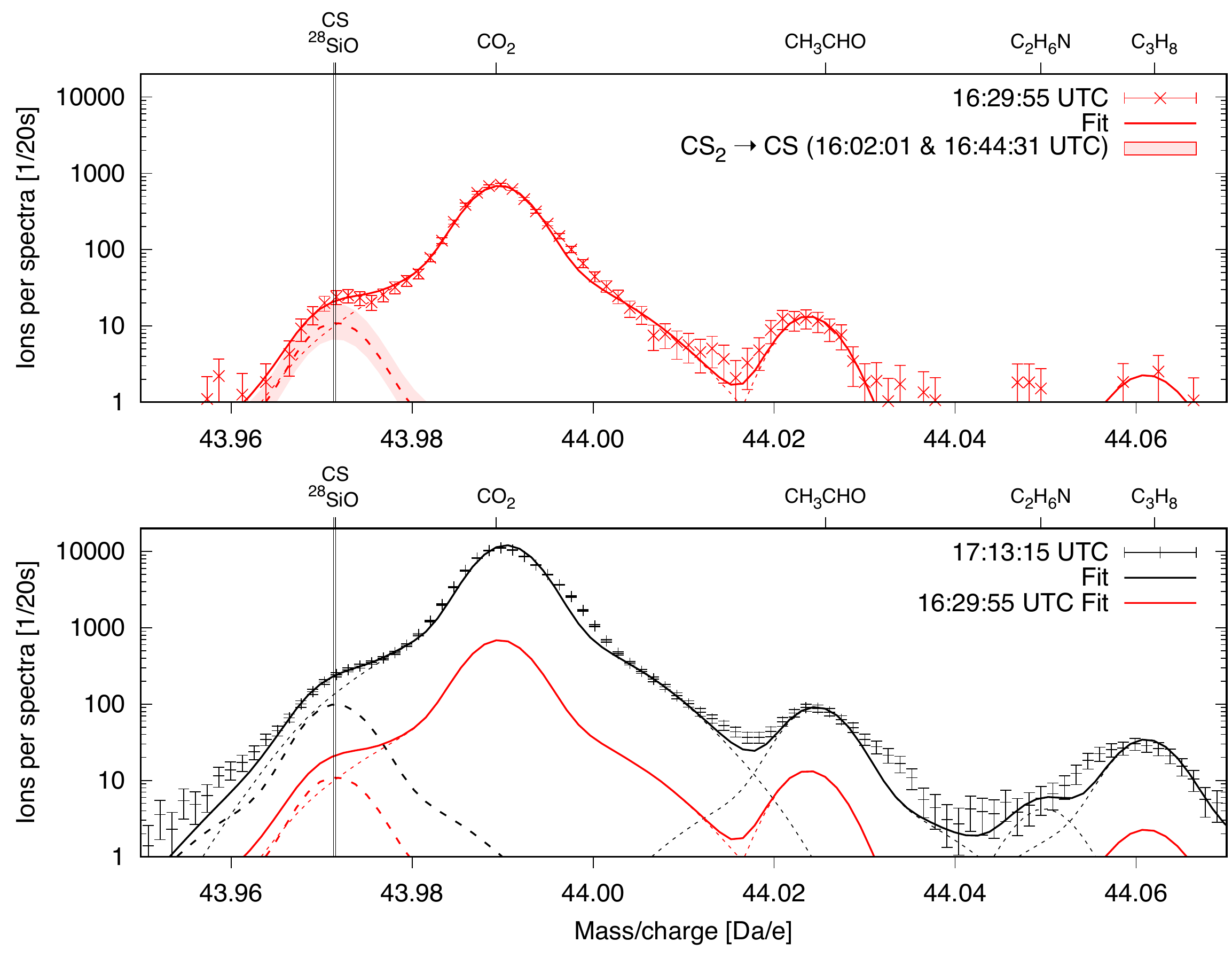}
   \caption{Two ROSINA DFMS m/Q~=~44~Da/e spectra top and bottom, measured on 31 July 2015 during and after the slew, 16:29:55 (MC\_20150731\_162827570\_M0222.TAB, row~B) and 17:13:15~UTC (MC\_20150731\_171147603\_M0222.TAB, row~B), at off-nadir angles 130 and 19~degrees, respectively. $^{28}$SiO and CS cannot be separated and have been fitted as a single peak indicated by the red dashed line (dotted lines: all other species, solid line: sum curve; see Section~\ref{sec:DFMS} for information on the peak shape). The contribution to CS by fragmentation of cometary parent CS$_2$~$\rightarrow$~CS is indicated by the red shaded area and may fully account for the measured signal. The bottom panel also contains the sum curve of the upper panel obtained during the slew for comparison.}
              \label{fig:m44m44}%
    \end{figure*}

Figure~\ref{fig:GreatCircleScan} shows the measured signal of a suite of species of different volatility for comparison. The selection includes either parent or fragment fluorine F, coma CN (contribution of HCN$\rightarrow$CN fragmentation subtracted), $^{28}$Si, Na, CO, and NH$_3$ as a function of time between 31 July and 2 August 2015. The black solid line represents the angle between the pointing direction of DFMS with respect to the center of mass of the comet. During two consecutive slews the spacecraft was hence pointing in the direction away from the nucleus for a time long enough to cover the measurements presented in Figs.~\ref{fig:m28m29m30}, \ref{fig:m40m42}, and \ref{fig:m44m44}. The spectra in Fig.~\ref{fig:m28m29m30} have been obtained during the second slew shown in Fig.~\ref{fig:GreatCircleScan}, when the signals of CN, CO, and NH$_3$ dropped considerably, while the signal of the lesser volatile species Na and Si remained more or less stable. We therefore looked for other Si-bearing fragments during that period, in particular SiC, SiN, SiO, Si$_2$, SiP, and SiS, which combine the major elements previously identified in the gas phase at \mbox{67P/C-G} \citep{LeRoy2015,Rubin2019} with Si.

Figure~\ref{fig:m40m42} shows two mass spectra at m/Q~=~40 and 42~Da/e. The typical CH-signals, namely, C$_3$H$_4$ and C$_3$H$_6$ \citep{Schuhmann2019a}, can  still be observed despite the substantial off-pointing with regard to the nucleus. However, no signal at the m/Q for SiC and SiN could be identified, which provides upper limits of approximately $^{28}$SiC$\le^{28}$Si/100 and $^{28}$SiN$\le^{28}$Si/100.

Figure~\ref{fig:m44m44} shows that the detection of SiO is complicated due to the almost identical mass of CS. According to \cite{Calmonte2016}, CS measured by DFMS did not originate from \mbox{67P/C-G}, but was formed inside the ion source of DFMS. Specifically, electron impact ionization produces a CS signal due to fragmentation of CS$_2$. To disentangle the combined (CS+SiO) peak (the sum of both species corresponds to the red dashed curve in the top panel) the CS portion is estimated from the abundance of CS$_2$ by taking into account the corresponding fragmentation pattern. The red highlighted area shows the expected contribution of the CS fragment to the total signal based on previous and subsequent measurements of CS$_2$. Hence, the fragmentation of CS$_2$ may easily account for the full signal of the combined (CS+SiO) peak. Furthermore, one may also compare this spectrum obtained during the slew with the one obtained right afterwards when the instrument was again pointing in the direction of the comet. This comparison is shown in the bottom panel of \ref{fig:m44m44} in black, with the fitted curves from the top panel added for reference. It is evident that the combined (CS+SiO) peak decreases by a factor of about nine, similarly to the common cometary volatiles such as CO$_2$ ($\sim$15) and C$_3$H$_8$ ($\sim$12). However, Si, as shown in Fig.~\ref{fig:GreatCircleScan}, remained almost constant during that time, hence, the combined (CS+SiO) peak did not correlate with Si. Both observations are evidence that the amount of SiO is very small even though deriving an upper limit is difficult. A conservative estimate would be $^{28}$SiO~$\le$~$^{28}$Si/10.

   \begin{figure*}
   \centering
   \includegraphics[width=0.7\textwidth]{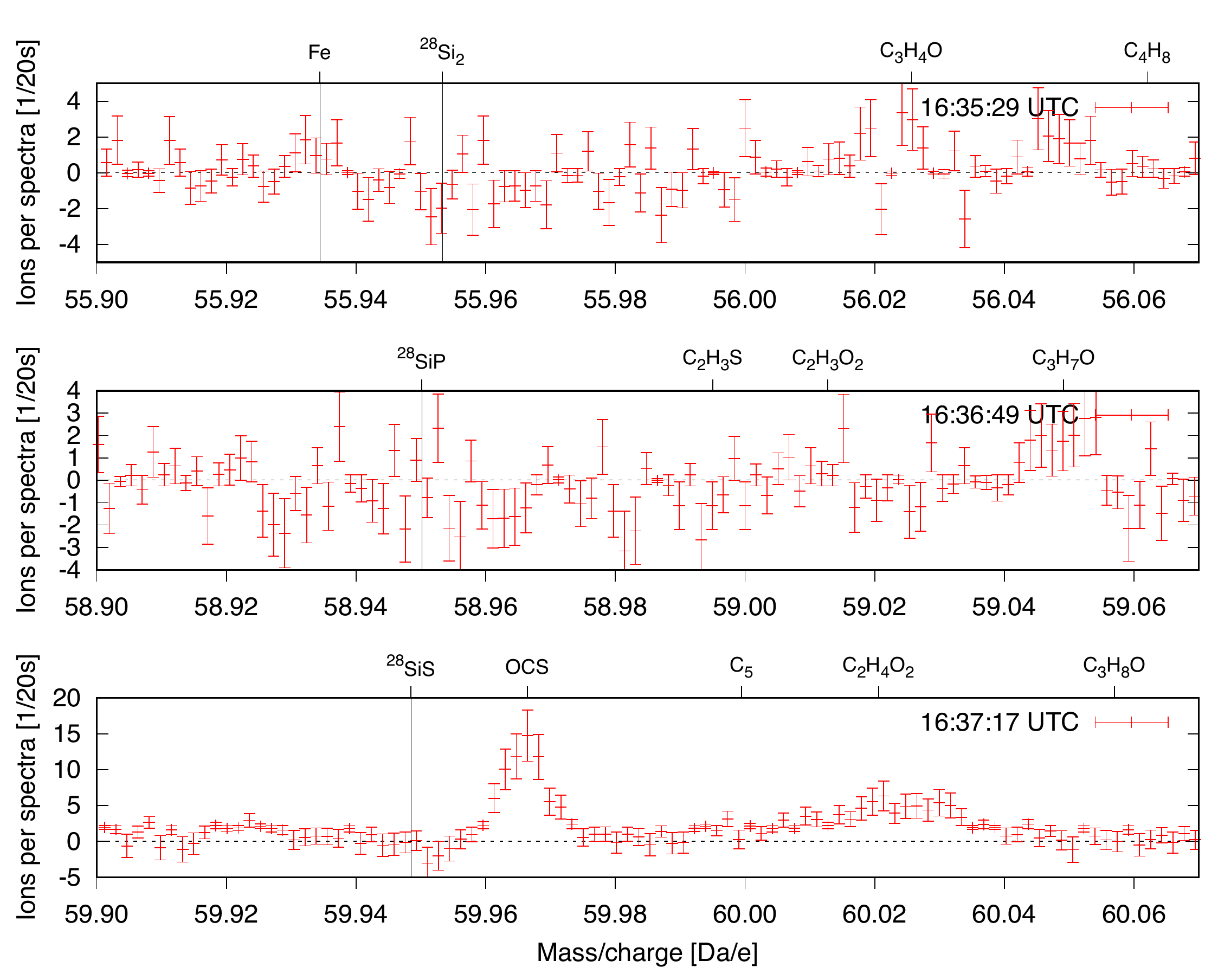}
   \caption{ROSINA DFMS m/Q~=~56, 59, and 60~Da/e spectra top to bottom, measured during the slew on 31 July 2015, 16:35:29 (MC\_20150731\_163401572\_M0222.TAB, row~B), 16:36:49 (MC\_20150731\_163521572\_M0222.TAB, row~B) to 16:37:17~UTC (MC\_20150731\_163549573\_M0222.TAB, row~B), at off-nadir angles between 115 and 110~degrees. The locations where Fe, $^{28}$Si$_2$, $^{28}$SiP, and $^{28}$SiS would be expected are indicated.}
              \label{fig:m56m59m60}%
    \end{figure*}

Similarly, Fig.~\ref{fig:m56m59m60} shows no significant signal of Si$_2$, SiP, or SiS on the corresponding m/Q~=~56, 59, and 60~Da/e, top to bottom. The analysis during the slew proved to be beneficial, especially for SiS, as the peak would be most of the time hidden under OCS, one of the major S-bearing species \citep{Calmonte2016}. These results lead to the corresponding upper limits of $^{28}$Si$_2$~$\le$~$^{28}$Si/100, $^{28}$SiP~$\le$~$^{28}$Si/100, and $^{28}$SiS~$\le$~$^{28}$Si/100, similar to previous results for the other species. 

   \begin{figure*}
   \centering
   \includegraphics[width=0.7\textwidth]{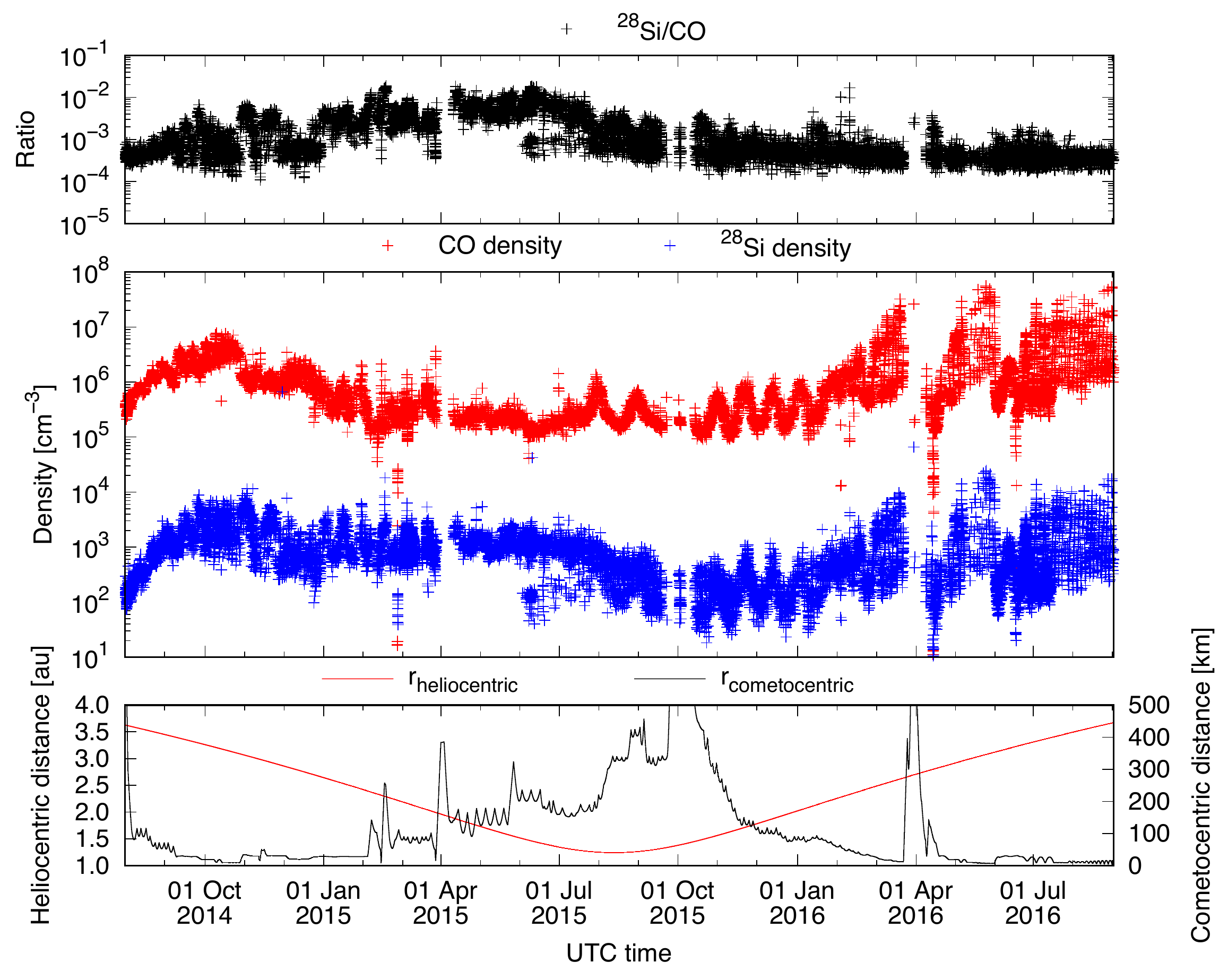}
   \caption{Timeline of the $^{28}$Si/CO ratio (top panel) and individual CO and $^{28}$Si densities (middle panel) measured in situ by ROSINA DFMS throughout the mission. The bottom panel shows the corresponding cometocentric distance of the Rosetta spacecraft and \mbox{67P/C-G's} heliocentric distance.}
              \label{fig:SiCOdensity}%
    \end{figure*}

By analyzing m/Q~=~28~Da/e spectra over the course of the mission we obtained the relative signal strengths of $^{28}$Si with respect to CO. Relating the signal of Si to CO instead of, e.g., H$_2$O has the advantage that both species were measured in the same spectrum, that is, at the same time (see Section~\ref{sec:DFMS}). First, we removed all fragment contributions to the CO signal; for instance, CO$_2$~+~e$^-$~$\rightarrow$~CO$^+$~+~2~e$^-$. Including the relative electron impact ionization cross-sections for Si \citep[$\sigma$~=~6.5~$\AA^2$; ][]{Freund1990} and for CO \citep[$\sigma$~=~2.036~$\AA^2$; ][]{Hwang1996} and the fragmentation ratio of CO into the parent ion with respect to all ionized fragments produced \citep[CO~+~$e^-$~$\rightarrow$~CO$^+$~+~2~$e^-$~=~0.964 with only negligible amounts of CO$^{++}$, C$^+$, and O$^+$; ][]{Rubin2019}, it is possible to derive a relative abundance of $^{28}$Si with respect to CO in terms of density at Rosetta. For this estimate, it was assumed that the instrument transmission is the same for both species, namely, that both $^{28}$Si and CO ions make it through the analyzer section in equal proportions. This is a reasonable approximation given the almost equal mass. It was also assumed that the detection yield on DFMS' MCP detector is the same for both species, which may introduce additional uncertainty. More details on obtaining relative abundances with DFMS can be found in \cite{Wurz2015}.

Combined with the densities of CO at Rosetta, available through ESA's Planetary Science Archive \citep[PSA; see][]{Besse2018}, the local densities of Si were estimated and the result is presented in Fig.~\ref{fig:SiCOdensity}. It shows the presence of Si throughout the time Rosetta escorted \mbox{67P/C-G}. Simultaneous variations of the CO and $^{28}$Si densities are observed when the spacecraft moved to denser regions in the coma. The Si/CO ratio varied over the course of the mission, with a low ratio early and late in the mission at larger heliocentric distances and around peak activity. However, it is not clear if the variation is caused by sputtering or the absence thereof; namely, if the low relative Si early and late in the mission is due to low solar wind fluxes at increased heliocentric distances and low relative Si around perihelion due to attenuation of the solar wind \citep{Behar2017}. The fact that Si and CO were correlated at certain times (e.g., late 2015-early 2016) and not correlated at other times (e.g., early November 2014) shows that the Si is of cometary origin and not associated with such effects as spacecraft outgassing or the decomposition of spacecraft materials \citep{Schlappi2010}.

In terms of average local gas density, we obtained a median of the ratio of $^{28}$Si/CO~$\sim$~7$\cdot$10$^{-4}$ but with significant variations throughout the mission (see Fig.~\ref{fig:SiCOdensity}). For the Si/H$_2$O~$\sim$~2$\cdot$10$^{-5}$ ratio we use this median in combination with CO/H$_2$O~=~3.1$\cdot$10$^{-2}$ from the pre-perihelion period \citep{Rubin2019} or CO/H$_2$O~=~3.0$\cdot$10$^{-2}$ time-integrated over the entire Rosetta mission \citep{Lauter2020} and we neglect the higher mass isotopes, $^{29}$Si and $^{30}$Si. This ratio has a substantial uncertainty of up to a factor of ten due to small number statistics and possible fractionation effects given the limited volatility of the relevant refractory species.

\subsection{Sodium (Na)}

   \begin{figure*}
   \centering
   \includegraphics[width=0.7\textwidth]{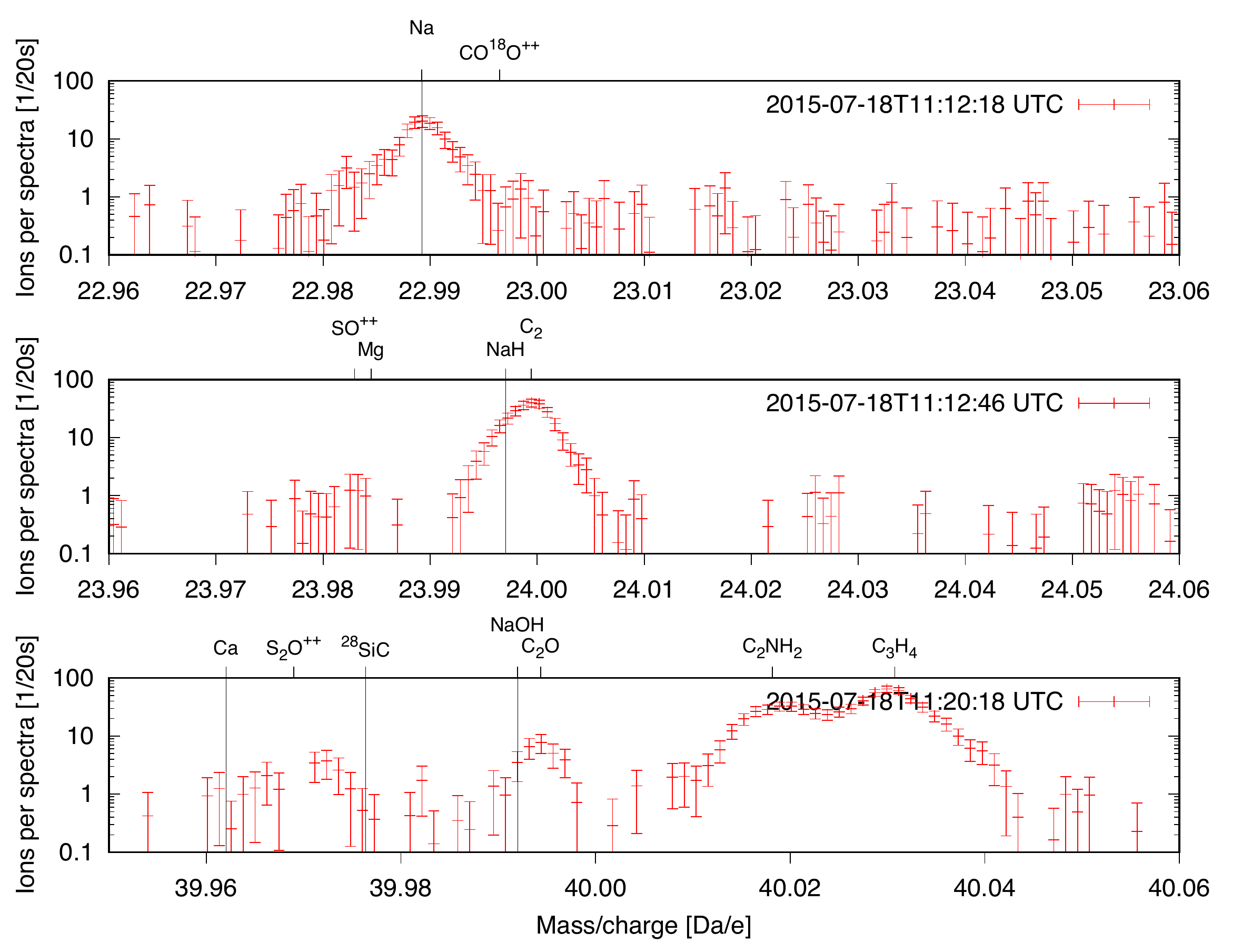}
   \caption{ROSINA DFMS m/Q~=~23, 24, and 40~Da/e spectra measured on 18 July 2015, 11:12:18 (MC\_20150718\_111051320\_M0222.TAB, row~B), 11:12:46 (MC\_20150718\_111119320\_M0222.TAB, row~B), and 11:20:18~UTC (MC\_20150718\_111851330\_M0222.TAB, row~B) top to bottom. Indicated are the masses of a suite of species including Na and where NaH and NaOH would be expected.}
              \label{fig:m23m24m40}%
    \end{figure*}

We also investigated m/Q~=~23~Da/e to search for the presence of Na, which had already been detected early in the mission when it most likely originated from sputtering \citep{Wurz2015}. The example spectrum in the top panel of Fig.~\ref{fig:m23m24m40} was obtained on 18 July 2015. Many more measurements showed the presence of Na close to the peak activity of the nucleus (see Fig.~\ref{fig:GreatCircleScan}), but count rates were generally low. Interferences at m/Q~=~23~Da/e were low also, occasionally showing a doubly-charged minor isotopologue of CO$_2$, that is, $^{12}$C$^{18}$O$^{16}$O$^{++}$. An analysis of the period from 26 to 29 July 2015 resulted in Na/Si~$\sim$~0.16 when using the electron impact ionization cross-section for Na \citep[$\sigma$~=~2.6~$\AA^2$; ][]{Fujii1995} relative to Si \citep[$\sigma$~=~6.5~$\AA^2$; ][]{Freund1990}  and the mass-dependent sensitivity correction according to \cite{Wurz2015}. Combined with Si/H$_2$O from Section~\ref{sec:Si} and neglecting minor Si isotopes follows Na/H$_2$O~$\sim$~3$\cdot$10$^{-6}$ by number. This ratio has again a substantial uncertainty of up to a factor of ten due to small number statistics and possible fractionation effects given the limited volatility of the involved refractory species.

The situation is somewhat more complicated at m/Q~=~24~Da/e, where we looked for the hydrated version of sodium, NaH (middle panel). Here, a strong interference with C$_2$ is found, a fragment of the numerous C-bearing species commonly found around \mbox{67P/C-G} \citep{Schuhmann2019a,Schuhmann2019b}. The observed C$_2$ peak may mask small amounts NaH; however, asymmetric peak shapes may occur at these masses (see also the peak shape of Na in the top panel for reference). Our upper limit for NaH/Na~$\le$~1 is therefore quite conservative. 

The bottom panel of Fig.~\ref{fig:m23m24m40} shows the m/Q~=~40~Da/e spectrum from the same measurement sequence. Here, we looked for NaOH, a possible parent for Na as suggested by \cite{Combi1997}; however, this search was unsuccessful. If NaOH is the major parent of Na and expelled by the nucleus, then it should be detectable by ROSINA despite the rather high photo-dissociation rate of 10$^{-3}$~s$^{-1}$ at 1~au \citep{Plane1991}. Taking an outgassing velocity of 1~km/s and correcting for the heliocentric distance of 1.28~au during this observation, Rosetta is within a fraction of the corresponding dissociation scale length of NaOH. Furthermore, the impact ionization by 45~eV electrons inside DFMS should form substantial amounts of parent ions. While no fragmentation patterns have been published, a ratio of Na$\le$NaOH/3 upon electron impact is estimated based on \cite{Huang2002}. However, at \mbox{67P/C-G} only a small amount of NaOH may be hidden under the left shoulder of C$_2$O, as shown in Fig.~\ref{fig:m23m24m40}. We obtain a conservative upper limit of NaOH$\le$Na/5. Hence the vast majority of Na in the coma of \mbox{67P/C-G} does not originate from NaOH. No information on the sublimation temperature of NaOH is available for the conditions at \mbox{67P/C-G}; however, the boiling point is very high \citep[$\sim$1660~K;][]{Haynes2011}, which would further complicate a cometary scenario involving NaOH. 

\cite{Combi1997} also investigated plasma sources for Na, but the issue remains the same: the close distance of Rosetta to \mbox{67P/C-G} reduces the time available for (photo)chemical reactions to occur to $\sim$100~s at a heliocentric distance of 1.25~au. 

\subsection{Iron (Fe)}

   \begin{figure*}
   \centering
   \includegraphics[width=0.7\textwidth]{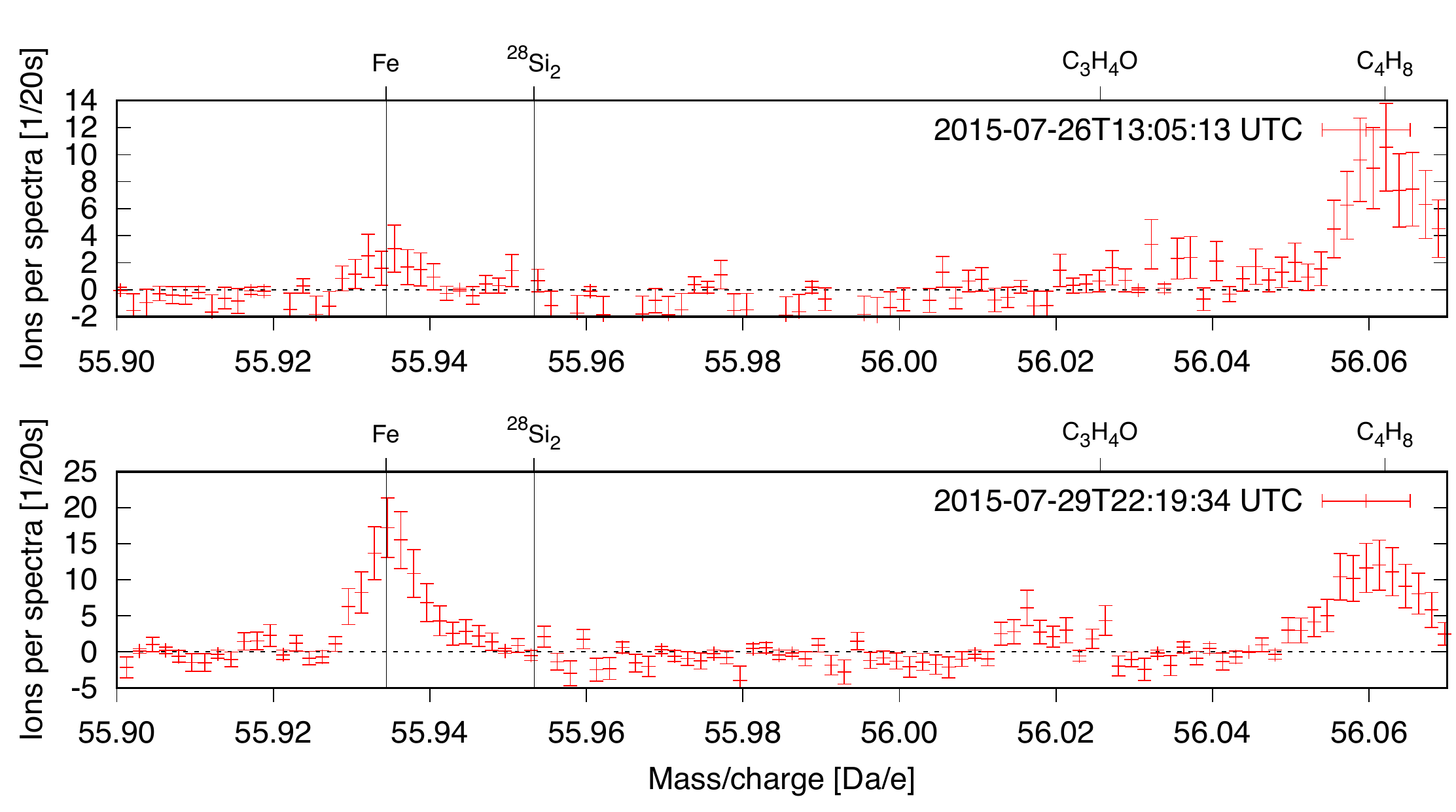}
   \caption{ROSINA DFMS m/Q~=~56~Da/e spectra measured end of July 2015 showing signals of Fe (top: MC\_20150726\_130345609\_M0222.TAB, row~B; bottom: MC\_20150729\_221806596\_M0222.TAB, row~A).}
              \label{fig:m56m56m56}%
    \end{figure*}

Based on earlier identification of iron and nickel atoms in a suite of comets \citep{Manfroid2021,Bromley2021} we searched DFMS m/Q~=~56~Da/e data for the presence of Fe. The spectra, one of row~A and one of row~B (cf. Section~\ref{sec:DFMS}), in Fig.~\ref{fig:m56m56m56} show strong evidence for the presence of Fe, more pronounced than in the top panel of Fig.~\ref{fig:m56m59m60}. Spectra with hints of Fe are very rare and the search for Ni at m/Q~=~58~Da/e was not successful. An analysis in the period from 26 to 29 July 2015 resulted in $^{56}$Fe/$^{28}$Si~$\sim$~2.1$\cdot$10$^{-3}$ using the electron impact ionization cross-section for Fe \citep[$\sigma$~=~5.0~$\AA^2$; ][]{Freund1990} relative to Si \citep[$\sigma$~=~6.5~$\AA^2$; ][]{Freund1990} and the mass-dependent sensitivity correction according to \cite{Wurz2015}. Taking this together with Si/H$_2$O (see Section~\ref{sec:Si}) and neglecting minor isotopes of Fe and Si leads to Fe/H$_2$O~$\sim$~5$\cdot$10$^{-8}$ in number. This ratio has again a substantial uncertainty of up to a factor of ten due to small number statistics and possible fractionation effects given the limited volatility of the involved refractory species.

   \begin{figure*}
   \centering
   \includegraphics[width=0.7\textwidth]{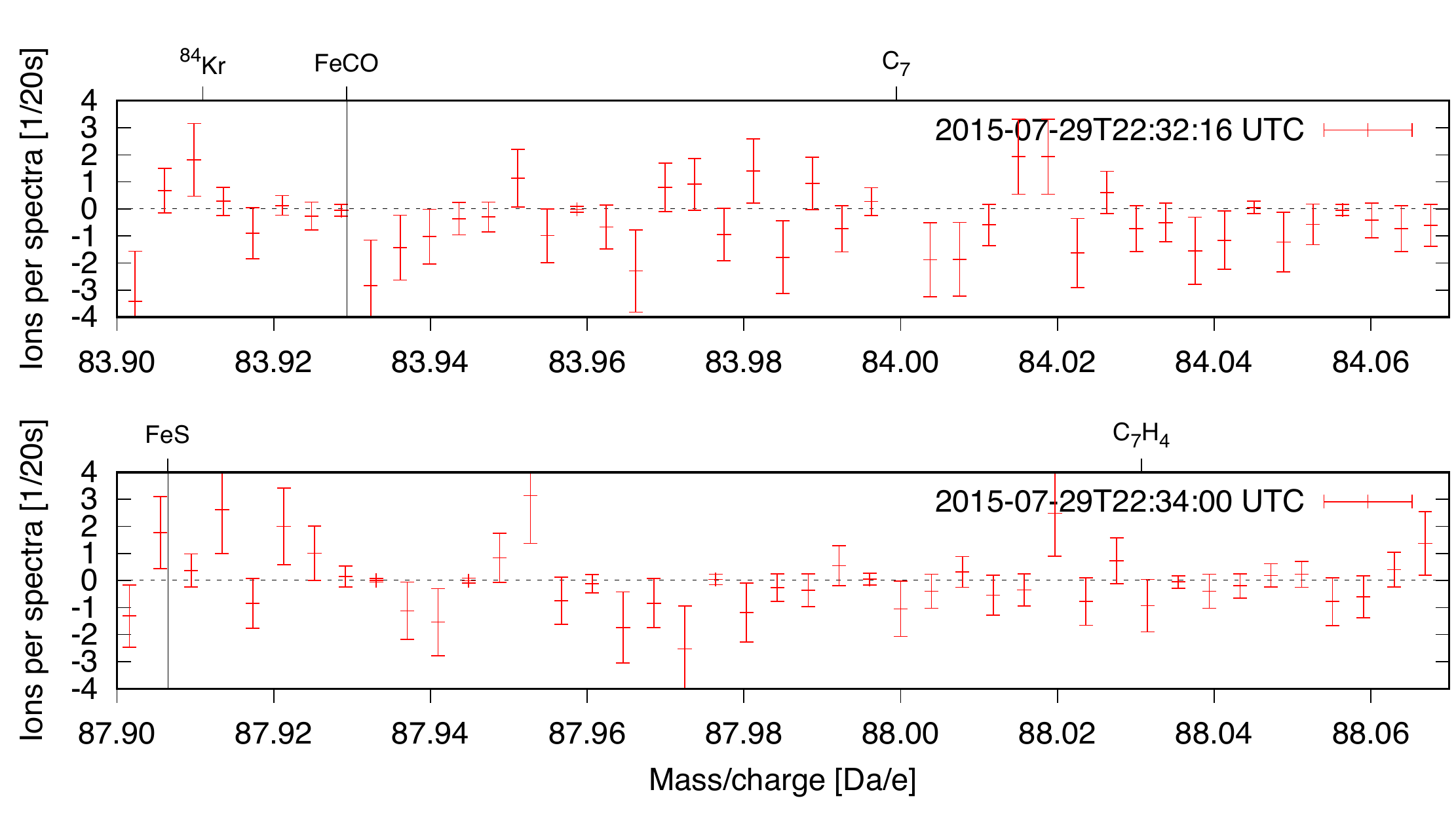}
   \caption{ROSINA DFMS m/Q~=~84 and 88~Da/e example spectra top and bottom, measured during the sequence where the Fe signal in the bottom panel of Fig.~\ref{fig:m56m56m56} has been obtained, i.e., on 29 July 2015, 22:32:16 (MC\_20150729\_223049289\_M0222.TAB, row~A) and 22:34:00~UTC (MC\_20150729\_223233306\_M0222.TAB, row~A), respectively. Indicated are the locations where FeCO, the major fragment of Fe(CO)$_5$ under electron impact, and FeS would be expected.}
              \label{fig:m84m88}%
    \end{figure*}

\cite{Manfroid2021} proposed short-living parent molecules for Fe, among them carbonyls including iron pentacarbonyl Fe(CO)$_5$, organometals, and other Fe- and Ni-bearing complexes. Many of these molecules are outside of the mass range of the measurement modes used during nominal DFMS observations. Furthermore, their volatility and abundance may have been too low for a detection. Taking into account electron impact ionization, from the list above, we looked for FeCO at m/Q~=~84~Da/e. This molecule is the major fragment of Fe(CO)$_5$ and formed about twice as much as Fe according to NIST \citep{Linstrom2001}.  Due to the employed post acceleration for atoms/molecules $\ge$70~Da/e, DFMS is almost equally sensitive to FeCO as to Fe (but again neglecting the species-dependent detector yield, as done for the estimation of the $^{28}$Si/CO ratio in Section \ref{sec:Si}). Nevertheless, as shown in the two example spectra in Fig.~\ref{fig:m84m88}, neither the detection of FeCO (top panel) nor troilite, FeS, which is found in meteorites (bottom panel), were successful. The same null result is obtained from the many other spectra we analyzed.

\section{\label{sec:discussion}Discussion}
In the previous section, a suite of detections and notable non-detections are reported.  Their relative abundances or upper limits are collected in Table~\ref{tab:ratios}.

\begin{center}
\begin{table}
\caption{\label{tab:ratios}Density ratios in the coma of \mbox{67P/C-G} at Rosetta with respect to H$_2$O. Uncertainties can be up to a factor of ten due to small number statistics and instrument fractionation.}
\begin{tabular}{ l c l }
 ratio & value & analyzed period\\ \hline
 \rule{0pt}{2ex}Si/H$_2$O &  $\sim$2$\cdot$10$^{-5}$ & entire mission, Fig.~\ref{fig:SiCOdensity} \\ 
 Na/H$_2$O & $\sim$3$\cdot$10$^{-6}$ & 26 - 29 July 2015 \\ 
 Fe/H$_2$O & $\sim$5$\cdot$10$^{-8}$ & 26 - 29 July 2015 \\ 
 Ni/H$_2$O & < 5$\cdot$10$^{-8}$ & 26 - 29 July 2015  \\ 
\end{tabular}
\end{table}
\end{center}

Table~\ref{tab:fragments} recaps the upper limits on fragments that may be formed through 45~eV electron impact from larger parent molecules inside DFMS. These limits exclude, for instance, silane SiH$_4$, sodium hydroxide NaOH, or iron pentacarbonyl Fe(CO)$_5$ from being present near Rosetta in amounts relevant to explain the measured Si, Na, and Fe, respectively.

\begin{center}
\begin{table}
\caption{Upper limits on fragments from 45~eV electron impact ionization of larger Si-, Na-, and Fe-bearing parent and daughter molecules present in the coma of \mbox{67P/C-G} near Rosetta. \label{tab:fragments}}
\begin{tabular}{ c l  lll}
 element & \multicolumn{3}{l}{max ratio} \\ \hline
  \rule{0pt}{2ex}Na & NaH & $\le$ & Na  \\
  & NaOH & $\le$ & Na/5  \\
 Si & SiH & $\le$ & Si/100 \\  
  & SiH$_2$ & $\le$ & Si/50 \\
  & SiC & $\le$ & Si/100 \\  
  & SiN & $\le$ & Si/100 \\  
  & SiO & $\le$ & Si/10 \\  
  & Si$_2$ & $\le$ & Si/100 \\  
  & SiP & $\le$ & Si/100 \\  
  & SiS & $\le$ & Si/100 \\  
 Fe & FeCO & $\le$ & Fe \\  
  & FeS & $\le$ & Fe   
\end{tabular}
\end{table}
\end{center}

In the following, these measurements are put into context of what we know about the presence of atoms that are made up of refractory elements in the cometary gas coma.

\cite{Manfroid2021} reported a nickel-to-iron ratio of Ni/Fe=0.23$\pm$0.04 in comet \mbox{103P/Hartley}. However, the average ratio of the entire sample of about 20 comets was higher with Ni/Fe=0.87$\pm$0.62, namely, almost at unity and more than a factor of ten greater than the proto-solar ratio of Ni/Fe=0.056$\pm$0.005 \citep{Lodders2021}. As stated by \cite{Hutsemekers2021}, neither the classification or origin of the comet played a role, except that the spread of the measured Ni/Fe ratio is significantly greater in JFCs than OCCs.

The occasional presence of Fe in DFMS mass spectra and Ni being below the detection limit indicates a Ni/Fe ratio $<$1 in \mbox{67P/C-G's} gas coma. \cite{Manfroid2021} argued that elevated Ni/Fe ratios may occur when a fraction of the iron, as opposed to nickel, is incorporated in silicates, which require temperatures in excess of 1000~K for sublimation. The same would apply if a fraction of the iron is present in its metallic form. However, Ni and the remaining Fe locked in sulfides would be expected to already sublimate at lower temperatures and, hence, an elevated Ni/Fe ratio is expected in the 600 to 1000~K temperature regime. Still, these are high temperatures, which require dust grains to be very small, namely, in the nanometer size range and to be at low heliocentric distances \citep{Lien1990}, so that they can reach temperatures of several hundred K. At \mbox{67P/C-G} hot grains with temperatures of up to 630~K have been observed during a dust outburst roughly one month after perihelion \citep{Bockelee2017}. On the other hand, the Ni/Fe ratio in the survey of about 20~comets by \cite{Manfroid2021} did not exhibit a statistically significant dependence on the heliocentric distance which contests this scenario. Another possibility could be high-temperature transients due to the stochastic heating of very small grains of tens to hundreds of atoms by single photon absorption \citep{Draine2001}. This is an important process, leading to infrared emission in the ISM; however, the situation may be different for comets in our solar system.

For comet \mbox{103P/Hartley}, \cite{Manfroid2021} reported minor amounts of iron by mass, namely, 1~g of Fe per 100~kg of H$_2$O, which translates into a Fe/H$_2$O ratio of $\sim$~3$\cdot$10$^{-6}$ by number. At \mbox{67P/C-G}, the signal of Fe in DFMS was limited to a few spectra and with Fe/H$_2$O~$\sim$~5$\cdot$10$^{-8}$ lower on average (see Table \ref{tab:ratios}). Furthermore, for \mbox{103P/Hartley} followed: Ni/H$_2$O~=~(Ni/Fe)~$\cdot$~(Fe/H$_2$O)~=~0.239~$\cdot$~3$\cdot$10$^{-6}$~=~7.2$\cdot$10$^{-7}$. At \mbox{67P/C-G}, Ni was below detection limit when assuming an Fe-like electron impact ionization cross-section, hence, Fe/H$_2$O$_\mathrm{67P/C-G}$~$<$~Fe/H$_2$O$_\mathrm{103P/Hartley}$ and Ni/H$_2$O$_\mathrm{67P/C-G}$~$<$~Ni/H$_2$O$_\mathrm{103P/Hartley}$. 

In the interstellar comet \mbox{2I/Borisov}, Ni was produced at a rate of 0.3\% by number with respect to CN and 0.002\% with respect to OH \citep{Guzik2021}. Assuming Ni/OH~$\sim$~Ni/H$_2$O leads to Ni/H$_2$O~$\sim$~2$\cdot$10$^{-5}$. This is similar to the relative abundance of Si in the mass spectra of DFMS at \mbox{67P/C-G} shown in Table~\ref{tab:ratios}, that is, Si/H$_2$O~$\sim$~2$\cdot$10$^{-5}$.

We still don't know the details of how these elements are released into the coma and there remains the question about the degree of fractionation occurring in DFMS, a neutral gas mass spectrometer designed for the measurement of volatile species that is not optimized for refractories. There is also the question of why Si is so commonly observed with ROSINA DFMS but not Fe (and Ni). \cite{Manfroid2021}, for instance, reported that they did not observe refractory elements lighter than Fe in their spectra. Nonetheless, for the purposes of the the following discussion, we compared gas-phase Si with the reported findings for Fe and Ni, bearing in mind the limitations of our measurement technique and the fact that we are comparing different elements.

\cite{Guzik2021} proposed a gaseous parent molecule for Ni with a short lifetime of $340^{+260}_{-200}$~s at 1~au. If we assume, for the sake of simplicity, an outgassing velocity on the order of 1~km/s, a scale length in the range of 140 to 600~km is obtained. Similarly, \cite{Hutsemekers2021} confirmed that Ni and Fe atoms are either released from the nucleus or from a short-lived parent. Our Si dataset from the end of July 2015 reveals no indication for a parent species, namely, it suggests that Si is most likely released in atomic form. During these observations, the spacecraft was at roughly 200~km from the nucleus which was 1.25~au from the Sun. Scaling the minimum Ni-parent lifetime \citep{Guzik2021} based on the larger heliocentric distance one obtains $\lambda_{Ni,~\mbox{67P/C-G}}$~=~140~s~$\cdot$~1~km/s~$\cdot$~1.25$^2$~$\sim$~220 km, comparable to Rosetta's cometocentric distance. When assuming a similar scale length for the Si-parent as for the Ni-parent still a substantial number density of the Si-parent should be present at Rosetta (i.e., $\sim$1/e). Therefore, the upper limits reported in Table~\ref{tab:fragments} remain considerable. Of course, our data do not exclude Si-parent molecules as such, but may require a much shorter lifetime, namely, several scale-lengths within the 200~km distance from Rosetta to the nucleus. However, a similar analysis at greater heliocentric distances and much closer to the nucleus yielded the same result, which would require an even shorter scale length. 

Figure~\ref{fig:GreatCircleScan} also shows that during the slews, the signals of NH$_3$ and coma CN, namely, the two species possibly associated with (ammonium) salts \citep{Hanni2019,Hanni2020,Altwegg2020}, did not correlate well with Si. Furthermore, no hints for refractory salts have been obtained. These are all indications that Si -- if it is not present in atomic form in cometary material -- it is at least released as such; for instance, from a distributed source of grains in the coma.

We also estimate how much Si is observed in the gas phase compared to the Si content in cometary grains. If we take $^{28}$Si/H$_2$O~$\sim$~2$\cdot$10$^{-5}$ for the gas phase and take into account that most gas phase oxygen is in H$_2$O, while neglecting the higher mass Si isotopes, we obtain Si$_{gas}$/O$_{gas}\sim$~2$\cdot$10$^{-5}$. Taking Si$_{dust}$/O$_{dust}\sim0.182$ from \cite{Bardyn2017} and O$_{dust}$/O$_{gas}\sim0.51$ from \cite{Rubin2019} for a dust-to-ice ratio (D/I) equal to 1 we obtain: 
\begin{displaymath}
\dfrac{\rm{Si}_{dust}}{\rm{Si}_{gas}}=\dfrac{\rm{Si}_{dust}}{\rm{O}_{dust}}~\dfrac{\rm{O}_{gas}}{\rm{Si}_{gas}}~\dfrac{\rm{O}_{dust}}{\rm{O}_{gas}}\sim5,000,
\end{displaymath}
that is, 1 in 5,000 Si atoms in the dust phase is released into the gas phase, or 1 in 15,000 if we base our calculations on D/I=3. At \mbox{67P/C-G}, the dust was reported to be rich in sodium \citep{Schulz2015,Bardyn2017}. The corresponding number for D/I=1 is $\rm{Na_{dust}}/\rm{Na_{gas}}\sim2500$  ($7,700$ for D/I=3) with substantial uncertainty (cf. Table~\ref{tab:ratios}). For comparison, \cite{Combi1997} estimated $\rm{Na_{dust}}/\rm{Na_{gas}}\sim100$ for comet \mbox{1P/Halley} on the basis of several assumptions and according uncertainty. For iron and D/I=1, what follows is  $\rm{Fe_{dust}}/\rm{Fe_{gas}}\sim540,000$ ($1,600,000$ for D/I=3).

The suite of refractory elements were also observed during the portion of the orbit, typically within 2~au from the Sun, when the solar wind did not have access to the surface of \mbox{67P/C-G} nor even the region surrounding the Rosetta spacecraft itself \citep{Behar2017}. This excludes solar wind sputtering of either the surface of the nucleus or the cometary dust grains surrounding Rosetta as a major contributor to the measured signal as opposed to the situation at larger heliocentric distances early in the Rosetta mission \citep{Wurz2015}. Furthermore, other sources involving energetic ions impinging on the surface of the nucleus or on dust grains near Rosetta, such as water pick-up ions, may be rather poor candidates due to their low fluxes and high variability \citep{Heritier2018}. However, refractory elements produced by sputtering in the outer coma and transported to the location of Rosetta cannot be excluded.

For the ISM, it was argued that nanometer-sized C-bearing grains could be responsible for the observed depletion of Si and other elements in the gas phase through accretion onto dust \citep{Weingartner1999}. Such carbonaceous nanograins with trapped refractory material may be incorporated into comets and subsequently release these elements from grains on the comet's surface or in the coma. One example may be heteroatom-bonded polycyclic aromatic hydrocarbons \citep[cf.][]{Bodewits2021}. The dust in \mbox{67P/C-G} is rich in carbon \citep{Bardyn2017} and contains volatiles \citep{Altwegg2017,Pestoni2021}.

The presence of positively and negatively charged nanometer-sized grains at \mbox{67P/C-G} have been reported by \cite{LLera2020} based on measurements of Ion Electron Sensor (IES), which was part of the Rosetta Plasma Consortium \citep{Burch2007}. These IES observations of the grains showed that their arrival direction did not come from the nucleus but was instead more scattered. This scattering may in part explain why Si did not exhibit the strong decrease in signal during the slews shown in Fig.~\ref{fig:GreatCircleScan}. Such grains may reach the vicinity of DFMS or even enter the ion source (depending on the energy, charge, and polarity) and release the Si, which is then detected in DFMS. The details of the process, however, remain unclear thus far.

Last, but not least, while the emissions of the sodium D-lines have been observed at numerous comets over a range of heliocentric distances, the origin of Na remains a puzzle. \cite{Combi1997} suggested a distributed source involving parent ions due to the bifurcated cross-tail structure of the extended Na tail in 1P/Halley, which did bear resemblance to 1P/Halley's H$_2$O$^+$ distribution in the tail. Alternative models based on a near-nucleus source with radiation pressure or the dust tail distribution did not match the observed abundance or size of the Na tail. In the near-nucleus region of 1P/Halley, according to \cite{Combi1997}, Na could be released either directly from the nucleus or from a very short-lived parent with a lifetime of $\ll$1000~s.

ROSINA DFMS observations at \mbox{67P/C-G} support the direct release of atomic Na, possibly similar to Si discussed above, from (nano)grains on the comet's surface or its surrounding coma. The fragmentation of such grains may also expose fresh surfaces and lead to additional release of such elements, which would further complicate the distribution of Na in the near-nucleus region and down the tail.
Earlier observations by \cite{Calmonte2016} also revealed substantial amounts of atomic sulfur (S) in the coma of \mbox{67P/C-G}, which complements a surprising suite of elements without any sign of larger gaseous parent molecules.

\section{Conclusions}
Based on our results, we have drawn the following conclusions:
\begin{itemize}
\item Metal atoms such as Fe have been observed by the Rosetta/ROSINA instrument in the coma of comet \mbox{67P/C-G}, confirming remote observations in a suite of other comets \citep{Manfroid2021,Bromley2021}. These observations were performed while the comet was active near its perihelion and we obtained Si/H$_2$O~$\sim$~2$\cdot$10$^{-5}$, Na/H$_2$O~$\sim$~3$\cdot$10$^{-6}$, and Fe/H$_2$O~$\sim$~5$\cdot$10$^{-8}$ with substantial uncertainties of up to a factor of ten, respectively. However, we found Ni to be below the detection limit.

\item The measurements revealed the presence of Si in the gas phase in the coma of comet \mbox{67P/C-G} over a wide range of heliocentric distances. These observations show that sputtering by solar wind particles on the nucleus or near Rosetta, as identified early in the Rosetta mission \citep{Wurz2015}, are unlikely the sole release process due to the collisional attenuation of the solar wind in the coma \citep{Behar2017}. However, based on our results, the transport of atoms sputtered from grains at increased cometocentric distances cannot be excluded.

\item No parent molecules, or fragments thereof, for either Si, Na, or Fe, have been identified in our data set. Given the proximity to the nucleus of approximately 200~km, these observations favor the scenario that such elements are released directly from (nano)grains, rather than from a distributed source of gaseous parent molecules.

\item Unlike common cometary volatiles, such as H$_2$O and CO$_2$, the Si signal remained stable when the spacecraft was off-pointing from the nucleus. Hence, (nano)grains in the coma are favored over grains on the nucleus' surface as source for the Si observed at the location of Rosetta.

\item The release occurs most likely in atomic form, similar to S \citep{Calmonte2016} and the noble gases Ar, Kr, and Xe \citep{Balsiger2015,Marty2017,Rubin2018}. Refractory-bearing parent molecules cannot be excluded, however, they would require scale lengths  of $\ll$200~km at a heliocentric distance of 1.25~au. 

\end{itemize}

The question remains how exactly these refractory atoms end up in the gas phase and further studies are required to approach a credible answer.

\begin{acknowledgements}
      MR was funded by the State of Bern and the Swiss National Science Foundation (200020\_182418).
\end{acknowledgements}

%
 \bibliographystyle{aa_bibtex-style} 
 \bibliography{Refractories} 
%

\end{document}